\documentclass[aip,onecolumn,preprint]{revtex4-1}
\usepackage[inline]{trackchanges} %for better track changes. finalnew option will compile document with changes incorporated.
\usepackage{amsmath}
\usepackage{graphicx}
\usepackage{color}
\graphicspath{{./figs_rev/}}
\makeatletter
\def\input@path{{./figs_rev/}}
\makeatother
\newcommand{\ig}[2]{\includegraphics[width = #1]{#2}}

\begin{document}
%% ------------------------------------------------------------------------ %%
%  Title
%
% (A title should be specific, informative, and brief. Use
% abbreviations only if they are defined in the abstract. Titles that
% start with general keywords then specific terms are optimized in
% searches)
%
%% ------------------------------------------------------------------------ %%

\title{Bursty magnetic reconnection at the Earth's magnetopause triggered by high-speed jets}

%% ------------------------------------------------------------------------ %%
%
%  AUTHORS AND AFFILIATIONS
%
%% ------------------------------------------------------------------------ %%

% Authors are individuals who have significantly contributed to the
% research and preparation of the article. Group authors are allowed, if
% each author in the group is separately identified in an appendix.)

% List authors by first name or initial followed by last name and
% separated by commas. Use \affil{} to number affiliations, and
% \thanks{} for author notes.
% Additional author notes should be indicated with \thanks{} (for
% example, for current addresses).

% Example: \authors{A. B. Author\affil{1}\thanks{Current address, Antartica}, B. C. Author\affil{2,3}, and D. E.
% Author\affil{3,4}\thanks{Also funded by Monsanto.}}

\author{J. Ng}
\email{jonng@umd.edu}
\affiliation{Department of Astronomy, University of Maryland, College Park, MD 20742, USA}
\affiliation{NASA Goddard Space Flight Center, Greenbelt, MD 20771, USA}

\author{L.-J. Chen}
\affiliation{NASA Goddard Space Flight Center, Greenbelt, MD 20771, USA}
\author{Y. A. Omelchenko}
\affiliation{Trinum Research Inc., San Diego, CA 92126, USA}
\affiliation{Space Science Institute, Boulder, CO 80301, USA}

\begin{abstract}
The impact of high-speed jets -- dynamic pressure enhancements in the magnetosheath -- on the Earth's magnetopause has been observed to trigger local magnetic reconnection. We perform a three-dimensional hybrid simulation  to study the magnetosheath and magnetopause under turbulent conditions  using a quasi-radial southward interplanetary magnetic field (IMF). In contrast to quasi-steady reconnection with a strong southward IMF, we show that after the impact of a jet on the magnetopause, the magnetopause moves inwards, the current sheet is compressed and intensified and signatures of local magnetic reconnection are observed,  showing similarities to spacecraft measurements.  
\end{abstract}

\maketitle

\section{Introduction}

Magnetic reconnection is a fundamental plasma process in which the topology of magnetic field lines changes, often accompanied by the conversion of magnetic energy to kinetic energy \cite{yamada:2010}. Reconnection is important in space and astrophysical environments, and has been the subject of observational, experimental and computational study  \cite{yamada:2010,howes:2018}. 
In the Earth's magnetosphere, reconnection at the magnetopause and in the magnetotail drives global magnetospheric dynamics \cite{dungey:1961}, leading to magnetic storms \cite{gonzalez:1999} and substorms \cite{angelopoulos:2008}.

Understanding how the plasma environment affects the nature of reconnection is an area of interest. Reconnection is affected by factors such as turbulence \cite{lazarian:2020}, background flows \cite{liu:2016} and driving of the plasma \cite{rosenberg:2015}. We focus on reconnection at the Earth's magnetopause, which depends on magnetosheath conditions \cite{koga:2019}.

Upstream of the quasi-parallel regions of the Earth's bow shock (where the angle between the  magnetic field and the shock normal is less than $45^\circ$), the presence of reflected ions gives rise to streaming instabilities which excite ultra-low frequency (ULF) waves \cite{gary:1991}, which are then convected towards the bow shock. As they approach the bow shock, they can grow in amplitude and interact nonlinearly, which can lead to the formation of structures such as Short Large Amplitude Magnetic Structures (SLAMS) \cite{schwartz:1992,chen:2021}, spontaneous hot flow anomalies \cite{zhang:2013} and cavitons \cite{kajdivc:2011}. These structures have the potential to affect the magnetosheath and magnetopause \cite{lin:2005, blancocano:2009,omidi:2016}.

Magnetosheath jets, localised regions of enhanced dynamic pressure and plasma flow, are mainly found downstream of the quasi-parallel shock, with low IMF cone angles providing favourable conditions for their observation (Ref.~\onlinecite{plaschke:2018}, \onlinecite{raptis:2020} and references therein).
The effect of jets on the magnetopause has been the subject of numerous studies. They can cause Earth-radius scale indentations at the magnetopause \cite{amata:2011, hietala:2012, archer:2012,escoubet:2020} and drive surface waves \cite{archer:2019}. 
It has also been shown that jets can cause magnetopause reconnection \cite{hietala:2018,han:2017}.  In Ref.~\onlinecite{han:2017} it is suggested that  reconnection triggered by jets may cause throat aurorae, while Ref.~\onlinecite{hietala:2018} provides \textit{in situ} observational evidence that  high-speed jet caused compression of the magnetopause boundary, triggering magnetopause reconnection.

Because of the importance of kinetic physics at the quasi-parallel shock, hybrid simulations (kinetic ion, fluid electrons) are an important tool for studying global processes that govern turbulent plasma dynamics in the foreshock and the magnetosheath. Compared to full particle simulations, the hybrid model captures ion kinetic effects at greatly reduced computational costs \cite{winske:2003}.

Hybrid simulations have had success in simulations of dayside processes. With respect to the quasi-parallel shock, simulations have shown the formation of kinetic structures such as  diamagnetic cavities \cite{lin:2005}, spontaneous hot-flow anomalies (SHFAs) \cite{omidi:2016,blancocano:2018} and their impact on the magnetosheath. The nature and effects of  magnetic reconnection at the magnetopause has also been studied under different southward IMF conditions \cite{wang:2019,hoilijoki:2017,hoilijoki:2019,pfaukempf:2020}. Under quasi-radial IMF conditions, two dimensional simulations have found jets and reconnection in the magnetosheath downstream of the quasi-parallel shock, but reconnection at the magnetopause was not studied \cite{karimabadi:2014}. More recently, three dimensional simulations have shown the formation of large high-speed jets and their impact on the magnetopause and the cusp\cite{omelchenko:2021jet}.

In this work, we use a three-dimensional hybrid  model to study the magnetosphere under quasi-radial IMF conditions. We focus on the interaction of  high-speed jets with the magnetopause and show that signatures of local reconnection are observed after jet impact, in a manner similar to observations \cite{hietala:2018}.

%
% The main text should start with an introduction. Except for short
% manuscripts (such as comments and replies), the text should be divided
% into sections, each with its own heading.

% Headings should be sentence fragments and do not begin with a
% lowercase letter or number. Examples of good headings are:

\section{Simulation setup}
\label{sec:sim}

In this study we use a space-time adaptive hybrid particle-in-cell simulation code, HYPERS \cite{omelchenko:2012,omelchenko:2021,omelchenko:2021jet}. Ions are evolved kinetically while the electrons are treated as a charge-neutralising fluid. The magnetic field is updated using Faraday's Law and the electric field is determined by a generalised Ohm's Law. 
The resistivity used in this model is as follows:
\begin{equation}
\begin{split}
    \eta = \eta_{ch} + \eta_{v} \\
    \eta_{ch} = \frac{4\pi\nu_{ch}}{\omega_{pe}^2}, \nu_{ch} = c_{ch}\omega_{pi}\left[1-\exp\left(-\frac{v_d}{3v_s}\right)\right]\\
    \eta_{v} = \frac{4\pi\nu_{v}}{\omega_{pi,sw}^2}, \nu_{v} = c_{v}\omega_{ci,sw}\exp\left(-\frac{2n_e}{n_{min}}\right)\\
    v_d = |\vec{J}|/en_e, v_s = \sqrt{\gamma T_e/m_i}.
\end{split}
\end{equation}

Here $\omega_{pe}$ and $\omega_{pi}$ are the electron and ion plasma frequencies respectively, $n_{min}$ is the cutoff (minimum) electron density allowed in the simulation, $\vec{J}$ is the current density, $T_e$ is the electron temperature and $m_i$ is the ion mass. $\omega_{ci,sw}$ and $\omega_{pi,sw}$ are the initial solar-wind values of the respective quantities. 
$\eta_{ch}$ is the Chodura resistivity, which is an empirical expression  previously used to model field-reversed configurations \cite{omelchenko:2015}. The ``vacuum resistivity'' $\eta_v$ is vanishingly small in cells occupied with plasma ($n_e \gg n_{min}$). 
The other parameters are chosen to be small enough not to significantly alter wave dispersion in the regions of interest: $c_{ch} = 0.01$, $c_v = 0.5$, $n_{min}=0.1 n_0$, where $n_0$ is the solar wind plasma electron number density.   

The hybrid simulation  uses data from the run described in Ref.~\onlinecite{omelchenko:2021jet} at different times, and has a physical domain of 716$d_i\times$1334$d_i\times$1334$d_i$, where $d_i$ is the ion inertial length in the solar wind. The upstream ratio between thermal and magnetic pressure for both ions and electrons is $\beta =  8\pi n_0 T_0/B_0^2 = 0.5$, and the ratio of the speed of light to the Alfv\'en speed is $c/v_A = 8000$. The IMF cone angle is $10^\circ$ (southward) and the Earth's magnetic dipole is tilted 11.5$^\circ$ sunward. The IMF clock angle is zero ($B_y = 0$). The computational domain is covered by a 300$\times$500$\times$500 cell stretched mesh, with a uniform patch around the Earth covered by 175$\times$300$\times$300 grid cells with a resolution of 1 cell per $d_i$. Outside the uniform patch, the mesh cells are stretched exponentially with a growth factor of 12 \cite{omelchenko:2021jet}. For reference, the location of the uniform central patch is marked by the dashed lines in Fig.~\ref{fig:full}.

Solar wind plasma is injected at the left $x$ boundary of the simulation with velocity $v_x = v_0$ and travels towards the magnetic dipole, which is centred at the right $x$ boundary. The initial Alfv\'en Mach number is 8 (i.~e.~$v_0/v_A = 8$). The dipole strength is scaled such that the nominal magnetopause standoff distance is approximately $100$ $d_i$, which is much larger than the minimum distance required to simulate an earthlike magnetosphere ($\approx 20 d_i$)\cite{omidi:2004}. The inner boundary is a hemisphere of radius $50$ $d_i$ with absorbing boundary conditions for particles and perfectly conducting boundaries for fields. We note that distances are scaled down for computational feasibility such that $R_E \sim 12 d_i$ rather than realistically $\sim 60 d_i$. Unless otherwise mentioned, in the rest of the paper, $\rho = e n_e$ and $|\vec{B}|$ are normalised by their upstream values, $\vec{v}$ is normalised by the upstream Alfv\'en speed and lengths are normalised by the ion inertial length $d_i$. The current density $\vec{J}$ is normalised by $e n_0 c$, and $\Omega_{ci}$ is the ion cyclotron frequency calculated using the IMF. 

\section{Results}
\label{sec:res}

In the following analysis, we use the Geocentric Solar Magnetospheric (GSM) coordinate system, where the $x$-axis points from the Earth towards the Sun, the $y$ axis is in the dawn-dusk direction and the $z$-axis completes the right hand system.

\begin{figure}
  \ig{5.075in}{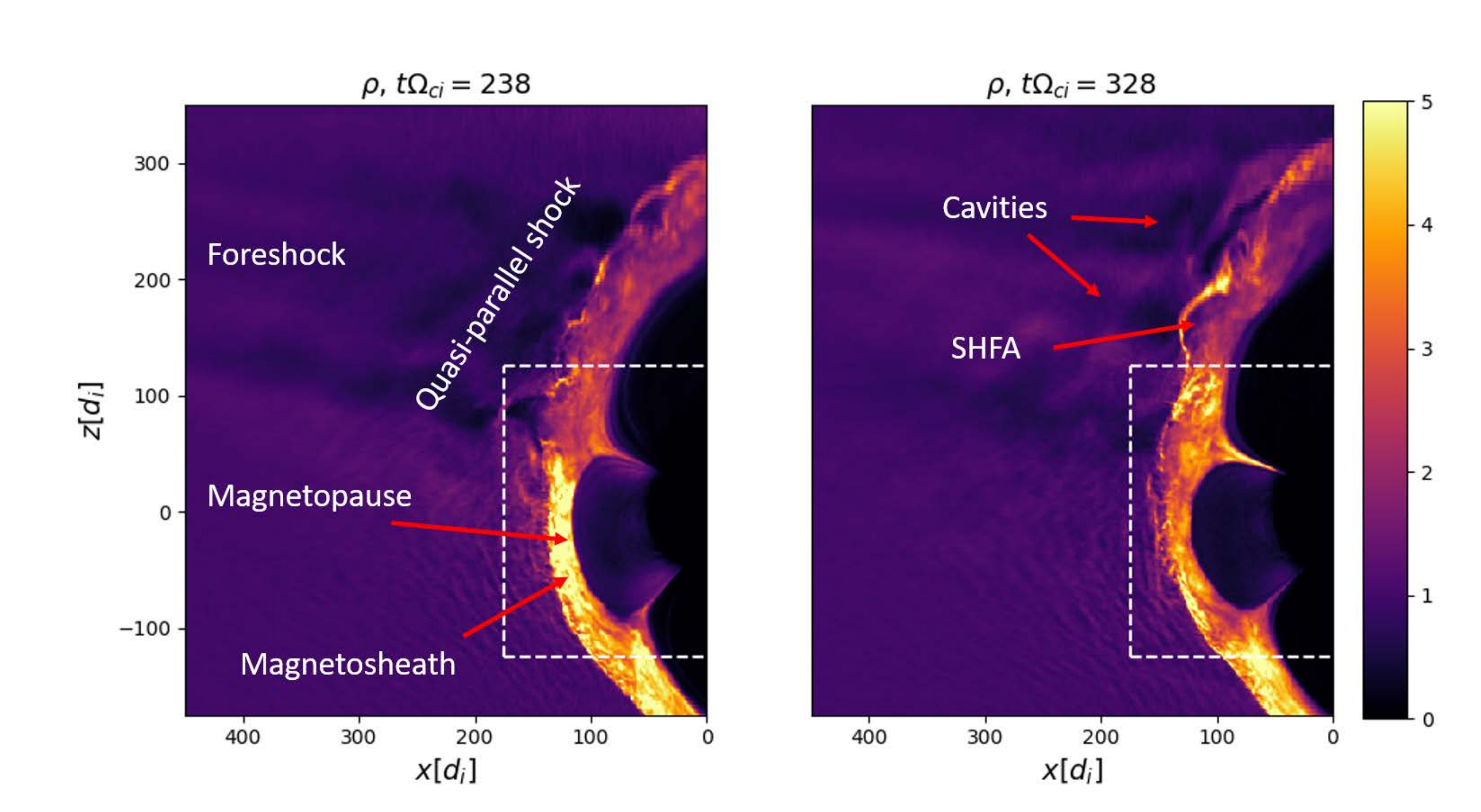}
  \caption{Overview showing the density in part of the simulation domain in the $x$-$z$ plane at $y = 0.5$, $t\Omega_{ci} = 238$ and $328$.  The dashed lines mark the boundary of the uniform central patch. }
  \label{fig:full}
\end{figure}

An overview of the system after the magnetosphere has developed is shown in Fig.~\ref{fig:full}. This shows the density in a subset of the $x$-$z$ midplane at $y = 0.5$. Upstream of the quasi-parallel shock there are density perturbations caused by the  waves in the foreshock, which are  convected towards the bow shock as described in other hybrid simulations \cite{omidi:2016, lin:2005, karimabadi:2014}. The quasi-parallel region of the bow shock is highly dynamic, with a rippled structure and no clear shock normal. Kinetic structures such as cavitons \cite{lin:2005,blancocano:2009} and spontaneous hot-flow anomalies (SHFA, one example is shown in Fig.~\ref{fig:full}) \cite{omidi:2016} are also  observed in this region as expected. As can be seen at the different times, the structure of the magnetosheath is dynamic in both space and time.

\begin{figure}
  \ig{4.275in}{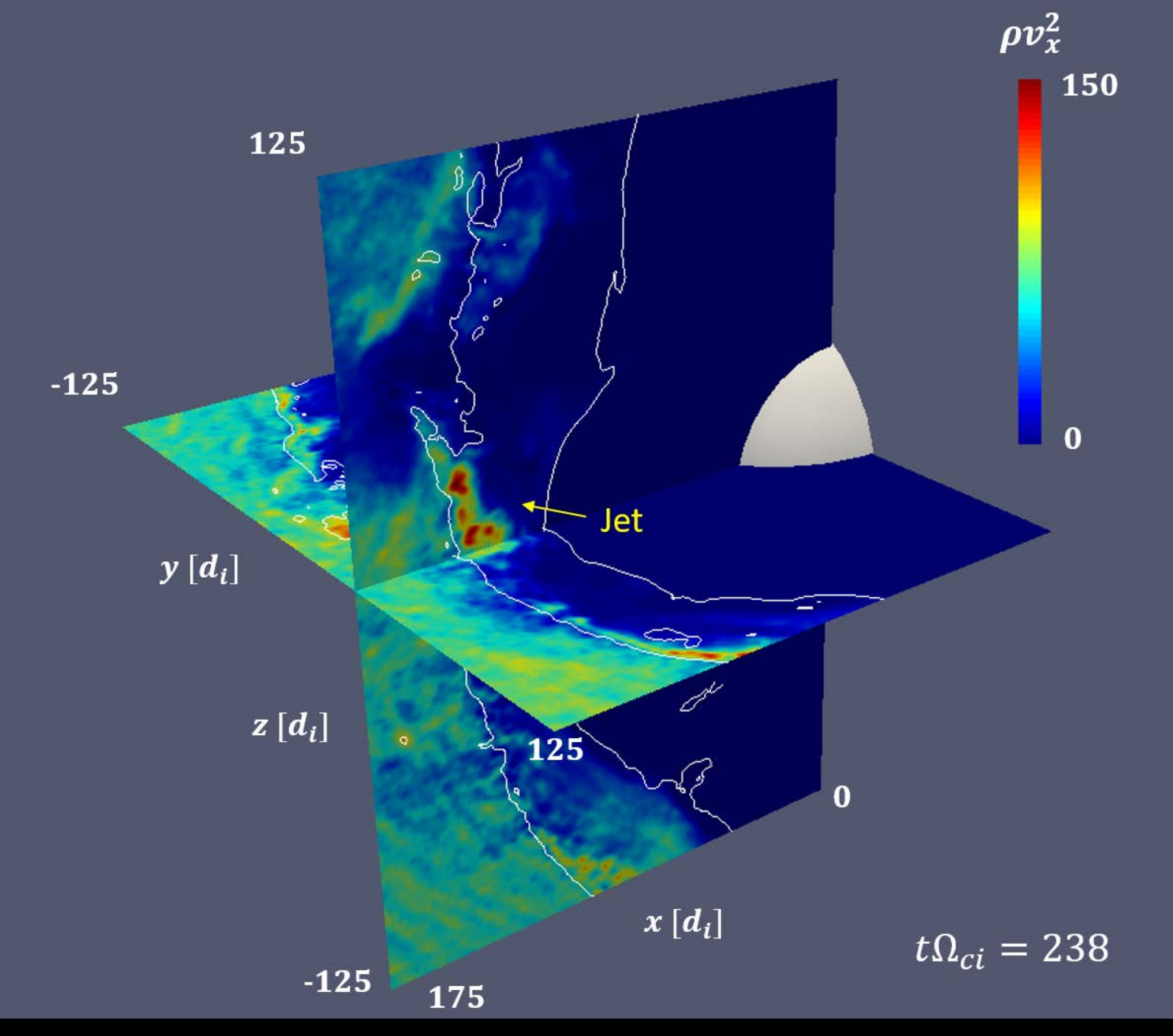}
  \caption{Slices showing the dynamic pressure $\rho v_x^2$ at $t\Omega_{ci}$ = $238$. White lines are contours of $\rho = 2$, giving the approximate locations of the bow shock and the magnetopause. The $x$-$z$ plane is at $y = 27.5$, while the $x$-$y$ plane is at $z = 10.5$. The region of enhanced dynamic pressure in the $x$-$z$ plane marks a magnetosheath jet. The hemisphere shows the simulation inner boundary.}
  \label{fig:3d_jet}
\end{figure}

The first time interval we study is $t\Omega_{ci} = 231$--$247$, during which a high-speed jet impacts the magnetopause, producing signatures of  magnetic reconnection. An illustration of the jet just before its arrival at the magnetopause is shown in Fig.~\ref{fig:3d_jet}, in which the dynamic pressure $\rho v_x^2$ at $t\Omega_{ci} = 238$ is plotted. Here a region of the enhanced dynamic pressure $\rho v_x^2$, which is used to identify the jet in the magnetosheath (e.g.~Refs.~\onlinecite{plaschke:2013,plaschke:2018,omelchenko:2021jet}), can be seen where the planes intersect. The approximate locations of the magnetopause and bow shock are marked by the white lines, which are contours of $\rho = 2$. The maximum dynamic pressure associated with the jet is approximately 180, compared to 64 for the solar wind.

\begin{figure}
  \ig{5.8in}{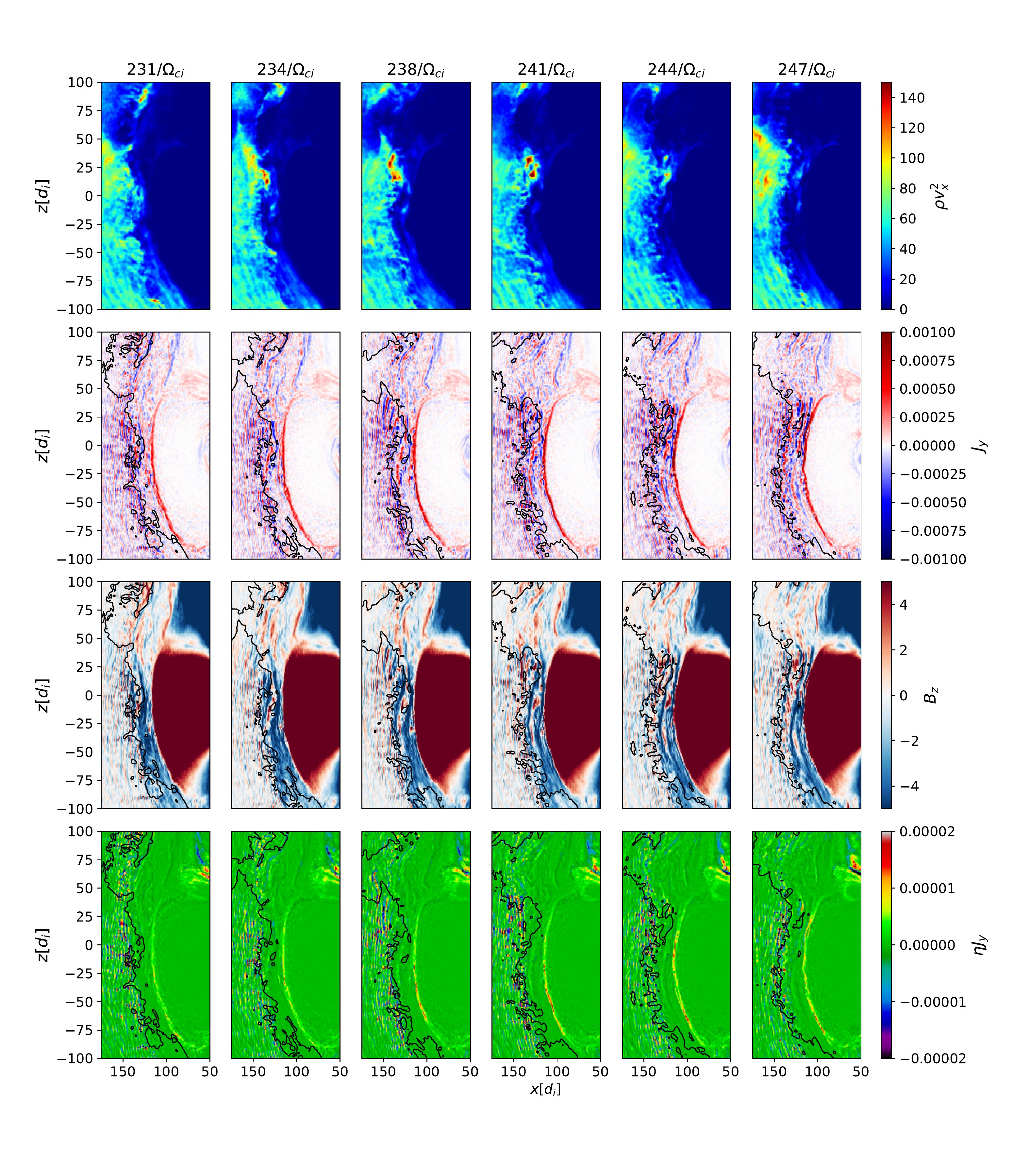}
  \caption{Evolution of the dynamic pressure $\rho v_x^2$, the current density $J_y$, the $z$-component of the magnetic field $B_z$ and the $y$-component of the resistive term in Ohm's law.  The black contours are  $\rho v_x^2 = 0.5 \rho_{0} v_{0}^2$, showing the jet position relative to the magnetopause.}
  \label{fig:evolution_79}
\end{figure}

The evolution of the jet and its effect on the magnetopause can be seen in Fig.~\ref{fig:evolution_79}, which shows the dynamic pressure $\rho v_x^2$ and current density $J_y$ at $y = 27.5$ during the time interval considered. The dynamic pressure  threshold value $\rho v_x^2 = 0.5 \rho_{0} v_{0}^2$  is used to highlight the jet location relative to the magnetopause, and is plotted as black contours in the lower panels. Here the jet can be seen in the region with $0 \lesssim z \lesssim 25$. Following the impact of the jet, there is an inward perturbation of the magnetopause and an intensification of the current density, as can be seen in the region with $z > 0$.

\subsection{Magnetopause reconnection}

During this time interval, signatures of magnetic reconnection are observed. This can be seen in  Fig.~\ref{fig:evolution_79}, in which the evolution of $B_z$ and $\eta J_y$ is shown. We note that finite $\eta J_y$ is a necessary but not sufficient condition for magnetopause reconnection to take place. As such, signatures of large $\eta J_y$ will be used in conjunction with other quantities to verify if reconnection is indeed taking place. 

Because of the southward IMF, there exists a quasi-steady reconnection region at the magnetopause, in addition to any reconnection regions triggered by jets. Throughout the simulation, these reconnection regions are found at lower values of $z$, though their exact locations and intensities vary with time. An example can be seen in Fig.~\ref{fig:evolution_79} at $t\Omega_{ci} = 238$ in the lower panel where there is enhanced $\eta J_y$ at $z \approx -50$, away from the region where the jet impacts the magnetopause. The southward displacement of the reconnection region from the subsolar point is due to the presence of IMF $B_x$ and the dipole tilt\cite{zhu:2015,hoilijoki:2014,trattner:2007}.

\begin{figure}
  \ig{5.8in}{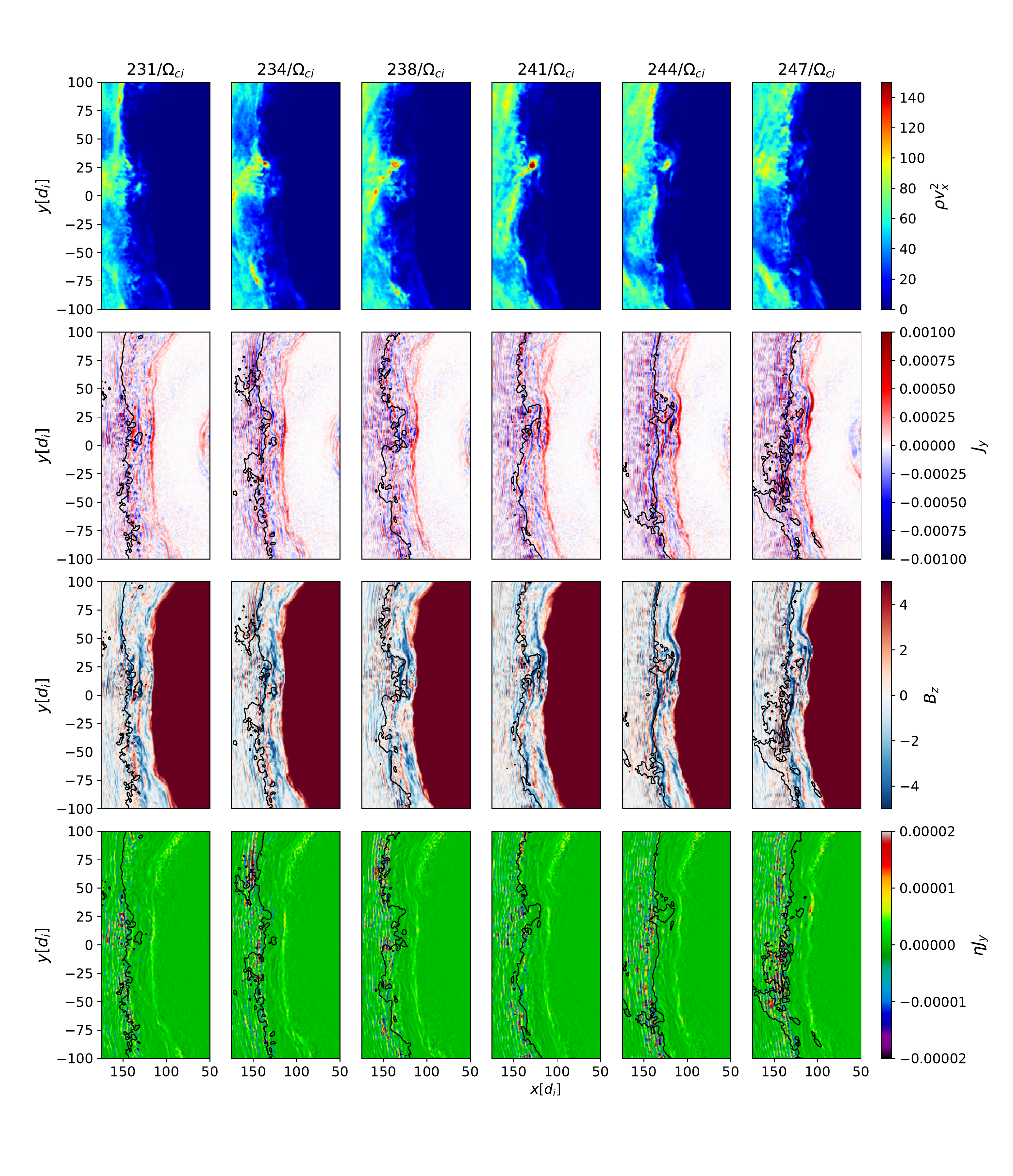}
  \caption{Evolution of the dynamic pressure $\rho v_x^2$, the current density $J_y$, the $z$-component of the magnetic field $B_z$ and the $y$-component of the resistive term in Ohm's law in the $x$-$y$ plane at $z = 18.5$}. 
  \label{fig:evolution_xy_79}
\end{figure}

After the arrival of the jet, another possible reconnection region with enhanced $\eta J_y\approx 2\times 10^{-5}$ develops at $z \approx 20$, close to the jet location shown in Fig.~\ref{fig:evolution_79}, with the highest intensity at $t\Omega_{ci} = 247$. The time evolution of $B_z$ is also consistent with this. At earlier times ($t\Omega_{ci} =231$-$238$), the field-reversal is narrowest in the region $-30 < z <-60$, giving rise to a more intense $d_i$ scale current sheet there with peak current $\sim 10^{-3}$, while the inward propagation and enhancement of negative $B_z$ leads to a stronger and thinner reversal in the $z > 0$ region at later times. In this region, the peak current density is doubled at the end of the interval compared to the beginning, while the field reversal layer thins from approximately 3 $d_i$ to 1.5 $d_i$. The reconnection signatures here last until $t\Omega_{ci} \approx 259$, corresponding to an interval of $15/\Omega_{ci}$, or approximately $90/\Omega_{ci,ms}$ using the magnetosheath ion cyclotron time. The structure in the $x$-$y$ plane is shown in Fig.~\ref{fig:evolution_xy_79}. In this series of  plots the indentation of the magnetopause after the jet arrives is visible in the region with $y > 0$, as is the local enhancement of $J_y$, negative $B_z$ and $\eta J_y$.

In the magnetosheath, $B_z$ (and $B_y$, not shown) has an oscillatory structure. As a result, the magnetosheath field lines become twisted due to the $B_y$ and $B_z$ fluctuations. At the lower $z \approx -50$ reconnection site, there is a small positive guide field $B_y \sim 0.2 |B_{0}|$, where $B_{0}$ is the value of the reconnecting magnetic field immediately upstream of the reconnection region in the magnetosheath. The x-line in the $z \approx 20$ reconnection region is not oriented in the $y$-direction but approximately $20^\circ$ from the negative $y$-axis in the +$z$ direction. This was calculated using the method of Ref.~\onlinecite{swisdak:2007}, and the result approximately bisects the angle between the upstream and downstream magnetic fields. There is a small guide field $B_g \sim 0.1 |B_0|$, mainly in the negative $y$ direction,

\begin{figure}
  \ig{5.8in}{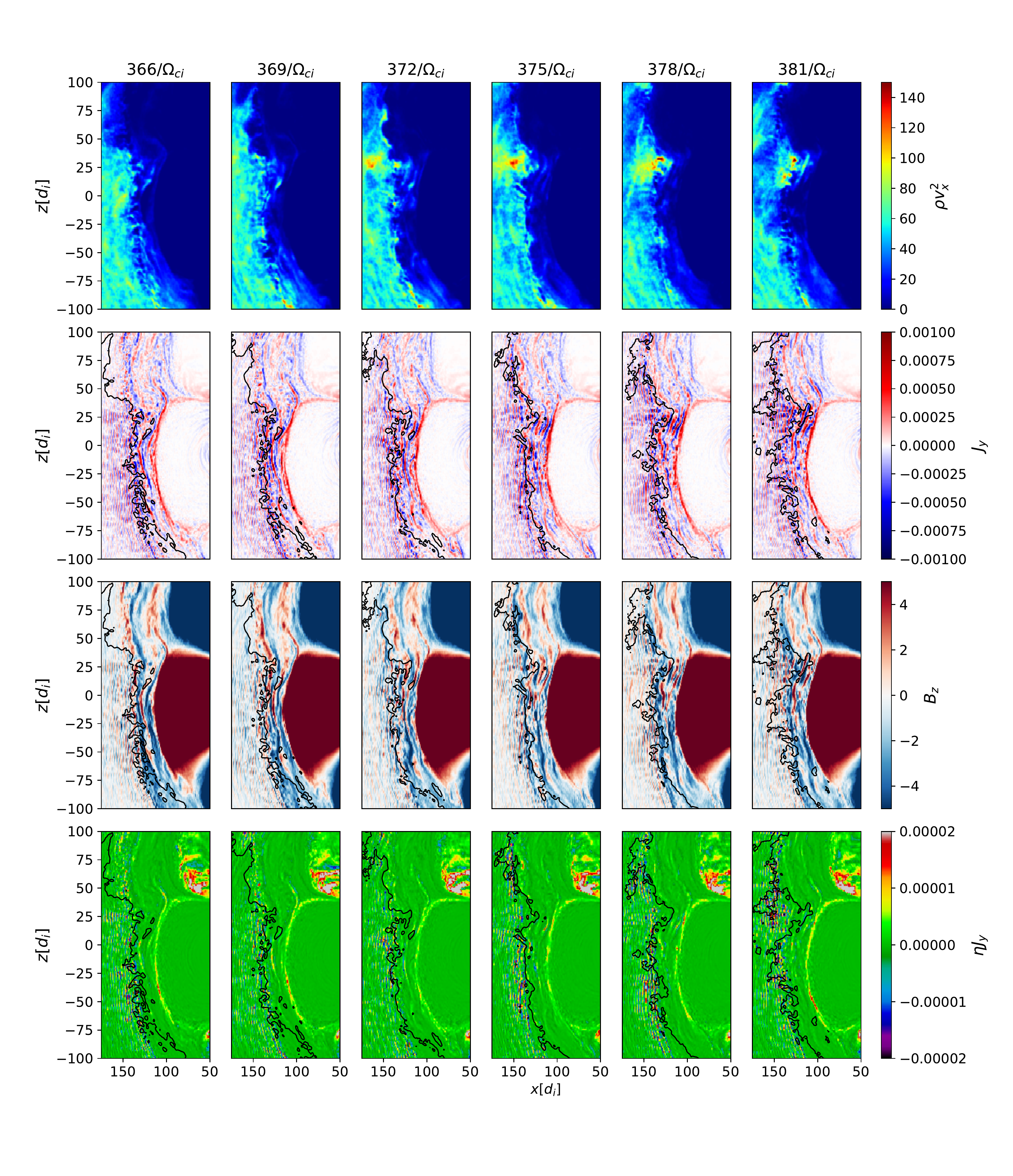}
  \caption{ Evolution of the dynamic pressure $\rho v_x^2$, the current density $J_y$, the $z$-component of the magnetic field $B_z$ and the $y$-component of the resistive term in Ohm's law in the plane $y = -32.5$. The black contours are  $\rho v_x^2 = 0.5 \rho_{0} v_{0}^2$, showing the jet position relative to the magnetopause. }
  \label{fig:evolution_119}
\end{figure}

In the simulation, the dayside magnetopause is intermittently bombarded by high-speed jets of different sizes and in varying locations. Jets substantially modify the ambient magnetic field. This creates conditions favourable for triggering reconnection events and we provide another example.  The evolution of the system during a second reconnection event is shown in Fig.~\ref{fig:evolution_119}. This takes place at a later time and in a different $y$ location at $y \approx -32.5$. As with the previous event, there is a quasi-steady reconnection region at negative $z$, and the arrival of a jet leads to the development of a reconnection region around $z = 20$. In this instance the x-line is oriented approximately $10^\circ$ from the negative $y$-axis in the +$z$ direction and there is a guide field  $B_g \sim 0.4 |B_0|$. The reconnection event here lasts until approximately $t\Omega_{ci} = 384$, for an interval of $6/\Omega_{ci}$ or approximately $30/\Omega_{ci,ms}$ (note that the upstream magnetosheath field is slightly weaker than in the previous event).  

To illustrate the reconnection geometry, we use the four-field junction method,  identifying where the field lines terminate and classifying them by their topologies (e.g.~Refs.~\onlinecite{laitinen:2006,pfaukempf:2020}). There are closed field lines which terminate at the inner boundary, open field lines which terminate at the domain boundary and half-open field lines which connect to both the domain and inner boundaries. Reconnection can occur where field lines with the four different topologies meet \cite{laitinen:2006}.

\begin{figure}
  \ig{6.375in}{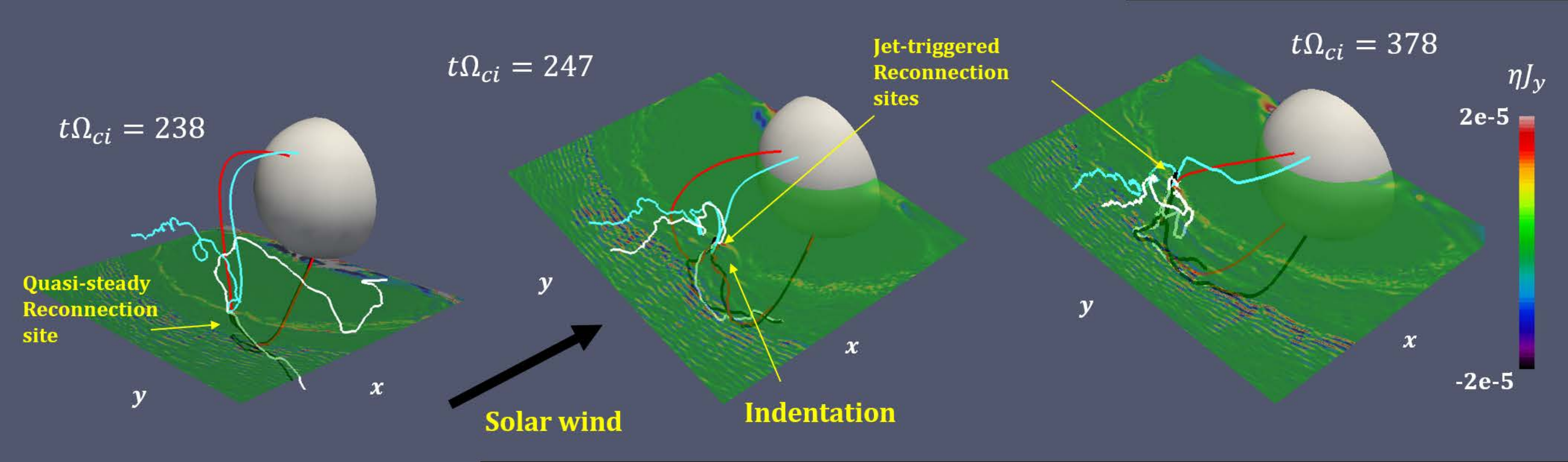}
  \caption{Three dimensional plots demonstrating quasi-steady and jet-triggered reconnection at the magnetopause. Slices show the extent of $\eta J_y$ at $t\Omega_{ci} = 238$, $247$ and $378$. The $x$-$y$ planes are at $z = -50.5$, $18.5$ and $21.5$ respectively. Magnetic field lines showing four different topologies are also shown in the region of enhanced $\eta J_y$.  The field lines are coloured by topology: the white lines start and end in the IMF; the black and cyan  lines are connected to the IMF and the southern and northern hemispheres respectively and the red lines are  closed. The line shading changes slightly when the lines pass behind the planes. The shaded domain corresponds to $x\in [0,175]$, $y \in [-125,125]$ in units of $d_i$.}
  \label{fig:lower_rx}
\end{figure}

In Fig.~\ref{fig:lower_rx} we show a three-dimensional view with slices of $\eta J_y$ at $t\Omega_{ci} = 238$, $247$ and $378$. There are field lines of four different topologies where $\eta J_y$ is enhanced, which is consistent with magnetic reconnection taking place. These figures illustrate the three different reconnection regions discussed earlier. At $t\Omega_{ci} = 247$, the inward dent in the magnetopause can also be seen in the $x$-$y$ plane, and has a width of approximately 30 $d_i$ ($\sim 2.5 R_E$).

\begin{figure}
  \ig{4.375in}{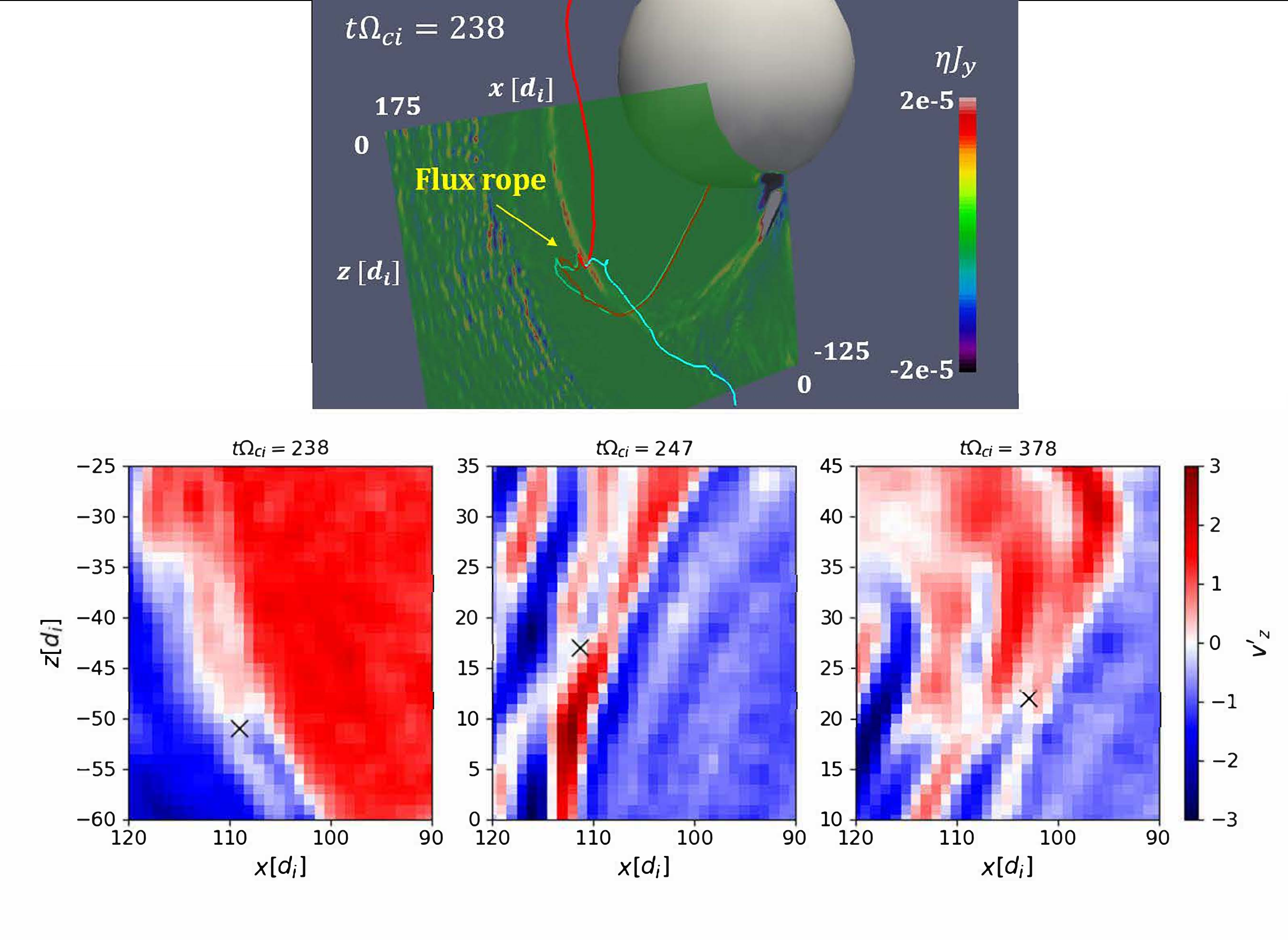}
  \caption{Top: Illustration of a flux rope at $t\Omega_{ci} = 238$ using the lower-half $x$-$z$ plane. Bottom: Ion velocity   $v'_z$  after subtracting the background flow at $t\Omega_{ci} = 238$, $247$ and $378$. The field junction points are marked by black crosses.  }
  \label{fig:vy}
\end{figure}

A flux rope, a signature of reconnection, is illustrated in Fig.~\ref{fig:vy} at $t\Omega_{ci} = 238$ approximately $5 d_i$ below the lower reconnection site. Here the twisting of the field lines is right-handed, consistent with expectations for southward IMF and positive $B_y$ \cite{kieokaew:2021}. Flux ropes are also seen at $t\Omega_{ci} = 378$. We do not observe flux ropes for the upper reconnection region at $t\Omega_{ci} = 247$.

\begin{figure}
  \ig{3.375in}{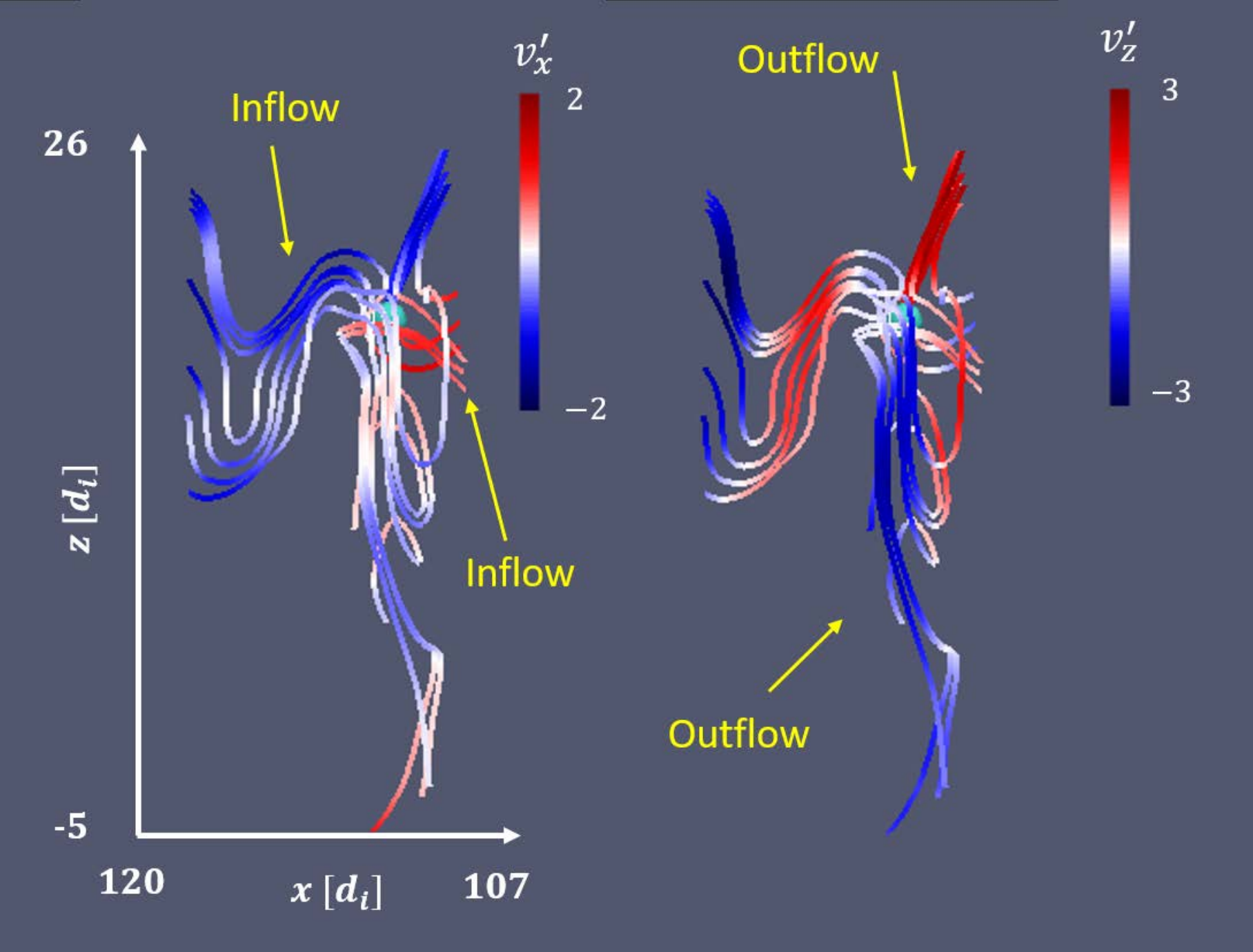}
  \caption{Three-dimensional streamlines of the ion velocity $\vec{v}' = \vec{v} - \langle \vec{v}\rangle$ in the vicinity of the reconnection region at $t\Omega_{ci} = 247$.  The cyan sphere marks the position of the field junction. Lines are shaded by $v'_x$ (left) and $v'_z$ (right). Viewing angle is along the $y$ axis. }
  \label{fig:3d_stream}
\end{figure}

Another diagnostic commonly used to identify reconnection regions is the ion outflow velocity. Although this has limitations due to the presence of background flows \cite{pfaukempf:2020}, it is illustrated in Fig.~\ref{fig:vy}, which shows the ion velocity $v_z$ at $t\Omega_{ci} = 238$, $247$ and $378$. Because of the background flow as well as fluctuations in the magnetosheath, the structure of the reconnection flows is complex. To improve their visibility, we calculate $v'_z = v_z - \langle v_z\rangle$, using the mean velocity in a $(10 d_i)^3$ volume around each junction. At $t\Omega_{ci} = 238$ and $378$, there are velocity reversals close to the field junctions, consistent with the presence of reconnection sites moving in the $-z$ and $+z$ directions respectively. The flow structure at $t\Omega_{ci} = 247$ is more complex and is illustrated in a three dimensional plot shown in Fig.~\ref{fig:3d_stream}. In this figure streamlines of the ion velocity in the background-flow frame are shown, and it can be seen that there is a flow divergence close to the field junction, which is marked by the cyan sphere. There are inflows in the $x$ direction and outflows in the positive and negative $z$ directions which are not  evident in the two-dimensional slices because these flows are not co-planar. The large negative values of $v'_x$ visible in the magnetosheath are due to the bulk flow, but the velocity is less negative immediately upstream to the reconnection region, with magnitude $< 0.1$ of the outflow speed.

In asymmetric reconnection, the predicted reconnection rate can be calculated using the formula \cite{cassak:2007}
\begin{equation}
  E\sim \frac{B_1 B_2}{B_1+B_2}\frac{v_{out}}{c}\frac{2 d}{L}
\end{equation}
where $v_{out}\sim B_1 B_2(B_1 + B_2)/(4\pi (\rho_1B_2 + \rho_2 B_1))$ is the asymmetric outflow speed, $B_{1,2}$ and $\rho_{1,2}$ are the upstream magnetic fields and densities respectively, and $d$ and $L$ are the half-width and half-length of the reconnection region. Due to the strong inhomogeneities in the magnetosheath, we use a spatially averaged magnetic field for the magnetosheath $B$-field values. When comparing the predicted results to the maximum $\eta J_y$, we find that the results are of the correct order, but smaller by approximate factors of $6$ for the $t\Omega_{ci} = 238$ event and $2.5$ for the $t\Omega_{ci} = 378$ event (or $\eta J_y/(B'v_{out}) \sim 0.013, 0.046$ respectively, $B' = 2 B_1 B_2/(B_1+B_2)$, consistent with the expected normalised reconnection rate of 0.01 to 0.1) . The discrepancies may be due to the inhomogeneities in the upstream parameters, and the measured outflow speeds being smaller than the model's asymmetric outflow used in the rate calculation.

%\begin{figure}
%  \ig{4.375in}{fig_edotj_para}
%  \caption{$\vec{E}\cdot\vec{J}$ at $t\Omega_{ci} = 238$, $247$ and $378$. Contours show $J_y = 0.0004$, highllighting the current sheet, and the field junctions are marked by the small spheres. The left plot shows the $x$-$z$ plane, which is approximately normal to the x-line. In the middle and right plots, the planes are normal to the x-line and are not parallel to the $x$-$z$ plane. The viewing angle is along the $y$ axis and the axis labels show the extent of the domain plotted in the $x$ and $z$ directions. }
%  \label{fig:edotj}
%\end{figure}

%\begin{figure}
%  \ig{3.375in}{lower_rx}
 % \caption{Three dimensional plot demonstrating reconnection at the magnetopause. Left: Two slices showing the 3D extent of $\eta J_z$ at $t\Omega_{ci}$ = $238$. Magnetic field lines showing four different topologies are also shown in the region of large $\eta J_z$. The $x$-$y$ plane is at $z=277.5$, while the $x$-$z$ plane is at $y = 199.5$. As in the previous figure, the field lines are coloured by topology: the white line starts and ends in the IMF; the black and cyan (behind the vertical plane) lines are connected to the IMF and the northern and southern hemispheres respectively and the red line is closed. The line shading changes slightly when the lines pass behind the planes. }
  %\label{fig:lower_rx}
%\end{figure}

\subsection{Magnetopause motion and magnetosheath fluctuations}

\begin{figure}
  \ig{3.375in}{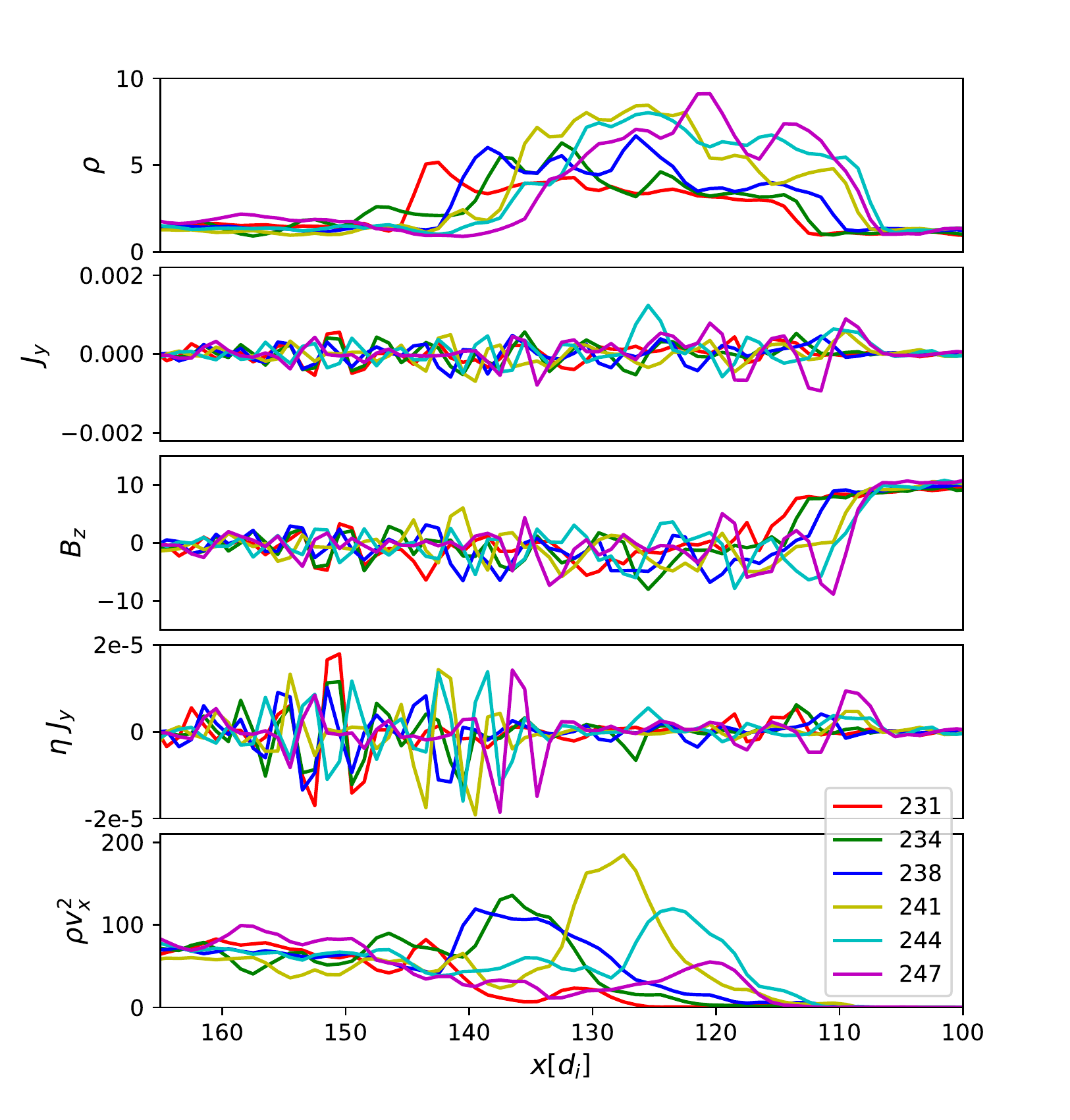}
  \caption{Cuts showing the one-dimensional temporal evolution of $\rho$, $J_y$, $B_z$, $\eta J_y$ and $\rho v_x^2$ at $y = 27.5, z= 18.5$. The inward perturbation of the magnetopause and the high-speed jet can be seen in the top and bottom panels respectively.   }
  \label{fig:line}
\end{figure}

A detailed view of the evolution of relevant physical quantities for the event at $t\Omega_{ci} = 247$ can be seen in Fig.~\ref{fig:line}. It shows the one-dimensional structure of the magnetosheath and magnetopause at $y = 27.5$, $z = 18.5$ where the jet-triggered reconnection site is. The inward perturbation of the magnetopause can be tracked using the density and $B_z$ plots, showing a displacement of approximately 6 $d_i$ ($\sim 0.5 R_E$). The approach of the jet towards the magnetopause can be seen in the bottom panel, which shows the evolution of the  dynamic pressure. Along this cut, the peak dynamic pressure is at $t\Omega_{ci} = 241$, but the jet still continues to approach the magnetopause after. This can also be seen in Fig.~\ref{fig:evolution_79}. The evolution is qualitatively similar during the second jet triggered event (not shown).

Reconnection signatures can also be seen in these traces. As $B_z$ evolves, the field reversal layer at the magnetopause becomes narrower, and the peak negative $B_z$ increases. Because of this, the current density and $\eta J_y$ at the magnetopause both increase, reaching maxima at $t\Omega_{ci} = 247$. We note that it is the thinning of the current sheet caused by the jet which leads to reconnection, in this case taking place after the inward motion of the magnetopause has stopped. 

The fluctuations with the oscillating $B_z$ structures may be related to the transmission of foreshock waves. This is similar to the propagating ``wavefronts'' seen in earlier two-dimensional hybrid simulations at the quasi-parallel region of the bow shock \cite{karimabadi:2014}. The transmission of large amplitude waves in quasi-parallel shocks has also been seen in observations at Venus \cite{shan:2014}. In this simulation, these fluctuations have a length scale of approximately $16 d_{i,ms}$, and frequency of approximately $0.2\Omega_{ci,ms}$, where we are using the local (magnetosheath) values of the inertial length and cyclotron frequency. The magnetic field fluctuations are primarily transverse ($B_y$ and $B_z$) and left-hand polarised in the simulation frame. Due to the inhomogeneous flows in the magnetosheath and the wave propagation velocity being similar to the bulk flow velocity, it is difficult to determine the polarisation in the plasma frame. 

%\begin{figure}
%  \ig{3.375in}{fig_line_119}
%  \caption{Cuts showing the one-dimensional temporal evolution of $\rho$, $J_y$, $B_z$, $\eta J_y$ and $\rho v_x^2$ at $y = 27.5, z= -32.5$. }
%  \label{fig:line_119}
%\end{figure}

\subsection{Discussion}

As mentioned in the introduction, there have been observations of high-speed jets causing $R_E$ scale indentations \cite{archer:2012,amata:2011,hietala:2012}, and triggering reconnection \cite{hietala:2018}. 
The events we study in our simulation show similarities with some features of these observations. The indentation we find at the magnetopause after the jet arrives has a scale size ($2.5 R_E$) similar to these observations. After the jets arrive, the field reversal region becomes narrower, and signatures of transient local reconnection are seen. These aspects are consistent with the observations. 

While there is magnetopause motion associated with the jet impact, the cause of reconnection in the observations was attributed to the compression of the current sheet, and reconnection was observed during the outward phase of the magnetopause motion \cite{hietala:2018}. In the simulation, the peak current density also occurs at the end of the time interval shown when the current sheet is thinner, after the jet arrives and the motion of the magnetopause begins to reverse, which can be seen most clearly in the density trace of Fig.~\ref{fig:line}.

%Earlier two-dimensional hybrid simulations have shown reconnection in the magnetosheath caused by the presence of jets causing field perturbations \cite{karimabadi:2014}. 

In the simulation there are additional effects  that may contribute to the transient magnetopause reconnection. As shown in Figs.~\ref{fig:evolution_79} and~\ref{fig:evolution_119}, during the same time  interval, a series of electromagnetic structures propagate towards the Earth. These structures bring enhanced negative (southward) $B_z$ to the magnetopause.  Similar electromagnetic structures have been found in other hybrid simulations (e.g Ref.~\onlinecite{karimabadi:2014}) and may be the result of the transmission and amplification of foreshock ULF waves \cite{shan:2014}. It is possible that the inward convection of the negative $B_z$ can lead to the narrowing of the field reversal regions and magnetic reconnection, as demonstrated in 3D hybrid simulations with quasi-radial northward IMF \cite{chen:2021jet}. However, for the events discussed earlier, the high-speed jets are clearly important for triggering the events. We use the $t\Omega_{ci} = 247$ event to further illustrate this point. In Fig.~\ref{fig:evolution_xy_79}, in which the $x$-$y$ plane is shown, the oscillatory $B_z$ structures have a large spatial extent in the $y$ direction. During and after the impact of the jet, the region of negative $B_z$ downstream of the jet shows the largest earthward displacement, leading to the steepening of the field gradient at the magnetopause and triggering  reconnection.

It should be noted that the jet-triggered reconnection events occur downstream of the quasi-parallel regions of the bow shock,  in regions of positive $z$ for the given dipole and IMF orientation. On the other hand, based on observations and other simulations\cite{zhu:2015,hoilijoki:2014,trattner:2007}, the locations of reconnection sites due to the geometry alone are expected to be displaced south of the subsolar point. This is confirmed in our simulations, in which quasi-steady reconnection events are found in the $z < 0$ regions. The jet-triggered reconnection events in our simulation, however are found  at $z > 0$, away from the typical reconnection regions. 

\section{Summary}
\label{sec:sum}

We have performed a 3D hybrid simulation with a steady quasi-radial southward IMF, focusing on establishing the relationship between magnetosheath jets and bursty reconnection at the magnetopause, in contrast to the quasi-steady reconnection commonly observed under strong southward IMF conditions. We find that  high-speed jets compress the magnetopause and reduce the width of the current layer. This leads to local magnetic reconnection with visible plasma outflows. The simulation results may explain the observations of jet-triggered magnetopause reconnection \cite{hietala:2018}. We also note the presence of strong negative $B_z$ structures convected towards the magnetopause which can potentially contribute to inducing  reconnection events.

%%%%%%%%%%%%%%%%%%%%%%%%%%%%%%%%
%% Optional Appendix goes here
%
% The \appendix command resets counters and redefines section heads
%
% After typing \appendix
%
%\section{Here Is Appendix Title}
% will show
% A: Here Is Appendix Title
%
%\appendix
%\section{Here is a sample appendix}

%%%%%%%%%%%%%%%%%%%%%%%%%%%%%%%%%%%%%%%%%%%%%%%%%%%%%%%%%%%%%%%%
%
% Optional Glossary, Notation or Acronym section goes here:
%

%%%%%%%%%%%%%%%%%%%%%%%%%%%%%%%%%%%%%%%%%%%%%%%%%%%%%%%%%%%%%%%%
%
%  ACKNOWLEDGMENTS
%
% The acknowledgments must list:
%
% >>>>	A statement that indicates to the reader where the data
% 	supporting the conclusions can be obtained (for example, in the
% 	references, tables, supporting information, and other databases).
%
% 	All funding sources related to this work from all authors
%
% 	Any real or perceived financial conflicts of interests for any
%	author
%
% 	Other affiliations for any author that may be perceived as
% 	having a conflict of interest with respect to the results of this
% 	paper.
%
%
% It is also the appropriate place to thank colleagues and other contributors.
% AGU does not normally allow dedications.

\acknowledgments
This work was supported in part by DOE grants DESC0016278, DESC0020058, NSF AGS-1619584, AGS-2010231, and NASA grants 80NSSC20K1312, 80NSSC18K1369 and 80NSSC19K0838. The authors thank Johnny Chang, Michael Heinsohn, Nancy Carney, and other NASA Advanced-Supercomputing and High-End-Computing team members for their professional support to make the simulation possible. The data that support the findings of this study are openly available in Zenodo at http://doi.org/0.5281/zenodo.5021233, Ref.~\onlinecite{ng:2021datajet3}.

%% ------------------------------------------------------------------------ %%
%% References and Citations

%%%%%%%%%%%%%%%%%%%%%%%%%%%%%%%%%%%%%%%%%%%%%%%
%
% \bibliography{<name of your .bib file>} don't specify the file extension
%
% don't specify bibliographystyle
%%%%%%%%%%%%%%%%%%%%%%%%%%%%%%%%%%%%%%%%%%%%%%%
\bibstyle{plain}
\bibliography{./reconnectionbib}

\end{document}